\documentclass[11pt,a4paper]{article}

\usepackage[margin=1in]{geometry}
\usepackage[utf8]{inputenc}
\usepackage[T1]{fontenc}
\usepackage{lmodern}
\usepackage{microtype}

\usepackage{amsmath,amssymb}
\usepackage{booktabs}
\usepackage{multirow}
\usepackage{array}
\usepackage{graphicx}
\usepackage{xcolor}
\usepackage[numbers,sort&compress]{natbib}
\usepackage{enumitem}
\usepackage{algorithm}
\usepackage{algpseudocode}
\usepackage{tcolorbox}
\tcbuselibrary{breakable}
\usepackage[colorlinks=true,linkcolor=blue!60!black,citecolor=blue!60!black,urlcolor=blue!60!black]{hyperref}

\newtcolorbox{llmprompt}{
    colback=gray!5,
    colframe=gray!40,
    boxrule=0.4pt,
    arc=1pt,
    left=6pt,
    right=6pt,
    top=6pt,
    bottom=6pt,
    fontupper=\small,
    breakable
}

\providecommand{\keywords}[1]{%
  \vspace{0.5em}\noindent\textbf{Keywords:} #1%
}

\title{\textbf{Active Learning of Approximate LLM Ranking for Energy-Efficient Retrieval of Long Queries}}

\author{
  Ohad Porat and Elad Yom-Tov \\
  \small Department of Computer Science and Artificial Intelligence \\ Bar Ilan University \\
  \small Ramat Gan, Israel
}

\date{}

\begin{document}

\maketitle

\begin{abstract}
Search engines and Large Language Models (LLMs) offer complementary strengths for document retrieval: lexical retrievers which rely on keyword matching are fast and energy-efficient, whereas LLMs capture semantic meaning and excel at interpreting long, context-rich queries. However, this comes at a cost that is computationally and energetically prohibitive. 

We propose the LLM Active Approximation Model (LAAM), a hybrid retrieval framework that introduces a lightweight learned predictor which serves as an intermediary between semantic retrieval and LLM retrieval. LAAM is initialized from a small set of LLM labeled documents and subsequently refined through an active learning loop which uses an LLM to label a small number of documents and incrementally update LAAM. By concentrating LLM evaluations on the most informative documents, the framework substantially reduces inference cost while preserving retrieval effectiveness.

We evaluate the proposed framework on TREC Robust 2004 using Lucene with BM25 retrieval and DeepSeek v3 as the LLM, under three levels of query verbosity. For verbose queries, LAAM achieves 87\% of average P@10 achieved by DeepSeek's reranking (0.459 vs. 0.526), while requiring only 3.3\% of the per-query energy consumption. LAAM is statistically significantly better than BM25 and is indistinguishable from DeepSeek on average precision (@1 \& @10), though not on MAP (Friedman test, $P<0.05$ with post-hoc testing). The resulting energy/accuracy tradeoff exhibits clear diminishing returns, indicating that most retrieval improvements are obtained with a small number of strategically selected LLM evaluations. In contrast, lower gains are exhibited for less verbose queries by both LAAM and the LLM. 

Overall, our results demonstrate that selectively approximating LLM relevance judgments through active learning enables retrieval accuracy to be controlled directly through the evaluation (and consequently energy) budget. Rather than committing to either inexpensive (and less accurate) lexical retrieval or exhaustive (and expensive) LLM reranking, our framework exposes a continuous spectrum of operating points along the energy--accuracy curve, allowing practitioners to select how to trade retrieval quality with computational budget. This establishes energy consumption as a first-class retrieval metric and offers a practical foundation for cost-aware retrieval over verbose information needs.

\end{abstract}

\section{Introduction}

Search engines serve as the primary method for finding information on the World Wide Web. They employ complex algorithms to collect, index, and rank web pages based on various factors, aiming to provide relevant results to user queries. LLMs are increasingly embedded inside the search stack. They are used for query and document expansion \cite{mao2020generation, nogueira2019document, wang2023query2doc}, serve as relevance rankers, where fine-tuned LLMs outperform traditional retrieval models \cite{ma2024fine, zhu2023large}, and can completely replace search engines and provide summaries of the relevant information. Current user studies reveal a complementary division of labor between the two technologies: search engines are preferred for quick factual lookups, while LLMs are favored for language-intensive queries that require contextual understanding \cite{caramancion2024large}.

The semantic gains of LLMs carry a steep cost in time and energy. Lexical retrieval answers in tens of milliseconds, whereas LLM inference can take several seconds per query \cite{bang2023gptcache, xiong2024search}. The energy gap is even larger. A single Google search emits an estimated 0.2 grams of CO$_2$ \cite{gombiner2011carbon} and consumes roughly 0.3~Wh, whereas a ChatGPT request has been estimated at 2.9~Wh, approximately ten times more \cite{devries2023growing}. At the document level, processing a single document with a BERT-based model has been estimated to emit approximately 5 grams of CO$_2$ \cite{gultekin2023energy}. Reranking hundreds of candidate documents per query with an LLM therefore multiplies the cost of retrieval by several orders of magnitude. This has motivated calls for ``green'' information retrieval research \cite{scells2022reduce}. Despite this, the trade-off between retrieval quality and computational cost is rarely treated as a first-order concern in retrieval system design.

This tension is sharpest for long queries. Typical web queries average only about 2.4 words \cite{bendersky2009analysis} and are thus often ambiguous \cite{ranpara2021contextual}. Verbose queries can disambiguate the information need, but they clash with the keyword-matching strategy of lexical retrieval \cite{gupta2015information}. LLMs, in contrast, are far better suited to interpreting such queries, owing to their ability to capture language semantics \cite{cheng2024potential}. As a result, the queries for which LLMs offer the greatest semantic benefit are those for which exhaustive LLM evaluation is most expensive.

Here we focus on the retrieval model, which matches user queries to relevant documents in the index. We present a hybrid retrieval component that exploits the advantages of keyword-based search engines (efficiency and low energy consumption) and of LLMs (accuracy) in different parts of the retrieval process so as to save energy while maximizing retrieval accuracy.

To achieve this, we propose the LLM Active Approximation Model (LAAM). In the proposed system, a query is first processed by a standard lexical retriever (e.g., BM25), which produces an initial ranked list of candidate documents. LAAM then acts as an intermediate layer between the lexical retriever and the LLM. A lightweight predictor learns to approximate the relevance judgments of the LLM from attributes of the query and the document, so that most documents never need to be evaluated by the LLM. Instead, a model initialized from a small set of LLM-labeled documents is refined through active learning, whereby at each iteration the system labels a single document using the LLM and updates the model. A final ranking that approximates LLM ranking is produced by the model, without having to label a large number of documents.

A key property of the proposed system is that the number of LLM evaluations is a controllable parameter, allowing a balance of accuracy with energy use. Rather than committing to either cheap lexical retrieval or costly full reranking, the system therefore allows practitioners to tune the balance between quality, energy use, and budget for long-query retrieval according to their deployment constraints.

We evaluate the framework on 50 TREC Robust 2004 \cite{voorhees2004overview} queries using Lucene with BM25 \cite{lucene, robertson1994some} and DeepSeek \cite{liu2024deepseek}, under three query formulations: title, description, and narrative. We show that for verbose narrative queries LAAM reaches approximately 87\% of the average P@10 achieved by full DeepSeek reranking (0.459 versus 0.526) while consuming only 3.3\% of the per-query energy, and the trade-off between energy and accuracy exhibits clear diminishing returns.

\section{Related Work}
In recent years, the integration of search engines with LLMs has garnered significant attention, especially with regards to retrieval models.
Retrieval models are at the heart of search engine architectures, responsible for identifying and ranking relevant documents in response to user queries. Over the years, various methodologies have been developed, ranging from early probabilistic approaches to models that leverage neural models \cite{zhu2023large}.
Search engines rely on indexing and ranking algorithms to retrieve relevant documents based on user queries. Most of the time, they work well with short queries \cite{silverstein1999analysis}. However, they often struggle with long queries that require nuances and context \cite{ranpara2021contextual}. In such situations, long queries can help disambiguate, but such queries do not work well with the search engines' keyword-matching strategy. On the other hand, neural model can capture semantics, enabling them to interpret meaning of long queries more effectively \cite{cheng2024potential}.

\subsection{Keyword-based and LLM-based retrieval}
Early search engines relied on term-matching. Vector space models (VSM) are a classical approach to document representation, where text documents are represented as vectors of words \cite{salton1975vector}. This was later improved through the Term Frequency-Inverse Document Frequency (TF-IDF) \cite{SALTON1988513} weighting scheme, which is a probabilistic method to estimate the importance of a word within a document relative to the entire collection.
Those Boolean retrieval models, which relied on exact matches of keywords, were gradually replaced by probabilistic models such as BM25 \cite{robertson1994some}, which introduced statistical approaches to rank documents by computing a nonlinear formulation of document properties.

With the rise of the web, search engines adopted more sophisticated techniques such as keyword-based search optimization and leveraging the web's link structure to rank retrieved results \cite{brin1998anatomy}.

LLMs enhance search engine retrieval beyond traditional algorithms. LLMs have been used for query expansion and pseudo-queries \cite{mao2020generation,wang2023query2doc} as well as document expansion \cite{nogueira2019document}. Ji et al. \cite{ji2022speeding} proposed using LLMs in two-phase method with offline keyword extraction and real-time query matching, improving efficiency and accuracy. By using an inverted index to narrow the search scope, their method substantially reduces the LLM's average response time while slightly improving answer quality. Feng et al. \cite{feng2023synergistic} introduce INTER, a framework that combines a retrieval module and an LLM to refine queries iteratively. In their design, the retrieval module expands query knowledge using LLM-generated collections while the LLM refines its prompts using the retrieved documents, leading to more accurate retrieval.

To address the limitations of LLMs, such as hallucinations and outdated data, Lei et al. \cite{lei2024corpus} developed Corpus-Steered Query Expansion (CSQE), merging LLM expansions with BM25-extracted key sentences to enhance relevance and retrieval effectiveness. These advancements demonstrate that LLMs significantly boost retrieval performance.

\subsection{LLMs vs. Search Engines}
A comparison of the strengths and weaknesses of LLMs and traditional search engines shows the advantages of each, and highlights the potential for developing hybrid models that leverage the strengths of both semantic search engines and LLMs.

\paragraph{User Preferences.}
User preferences regarding search engines or LLMs are influenced by numerous elements, including the kind of information desired, the length, context, and complexity of the query, and the user's expertise or comfort level with the technology \cite{caramancion2022interdisciplinary}.
Analyzing these preferences entails evaluating user satisfaction based on the tools' speed, accuracy, and overall user experience.
Caramancion et al. \cite{caramancion2024large} compared user preferences of search engines and LLMs in 20 different scenarios using a survey of 100 participants based in the US. The findings show that search engines are preferred for factual queries due to their breadth of indexed information and quick access to multiple sources, while LLMs are favored for subjective tasks and language-intensive queries due to their conversational and contextual understanding capabilities.

\paragraph{Quality.}
LLMs and search engines differ in the quality of retrieval, especially for complex and long queries: lexical search engines struggle to interpret verbose queries whose terms do not contribute equally to the information need \cite{gupta2015information}, while LLMs can follow the nuanced relevance criteria expressed in such queries and better capture their semantics \cite{weller2025followir,cheng2024potential}.
Sun et al. \cite{sun2023chatgpt} demonstrated that, when appropriately instructed, LLMs can act as effective zero-shot relevance rerankers, delivering competitive and in some cases superior ranking quality compared to state-of-the-art supervised retrieval models on standard IR benchmarks.
Ma et al. \cite{ma2024fine} showed that fine-tuned LLMs for retrieval outperform traditional retrieval models, demonstrating stronger generalization while directly processing long documents. This capability eliminates the need for the heuristic segmentation and pooling strategies commonly employed by conventional search engines to rank long documents.
The aforementioned examples provide evidence that LLMs add a layer of quality to the search engine system.

\paragraph{Time Cost.}
Xiong et al.~\cite{xiong2024search} noted that search engines excel in rapid indexing and retrieval, providing immediate access to information for fact-based queries, while LLMs take longer due to computational demands. Search engines typically deliver results within 10--100 milliseconds, while LLMs take 1--10 seconds or more per query. For example, OpenAI ChatGPT has an average response latency of 3 seconds \cite{bang2023gptcache}.
The cost of LLMs and of search engines can also be measured by their carbon footprint. Gultekin et al. \cite{gultekin2023energy} found that BERT-based models consume approximately 5 grams of CO\textsubscript{2} per document, while Gombiner et al.~\cite{gombiner2011carbon} found that a Google search query generates 0.2 grams of CO\textsubscript{2}.

\subsection{Active learning with LLM}
Active learning \cite{sammut2011encyclopedia} with LLMs is particularly beneficial in scenarios where labeled data is scarce or expensive to obtain, as it optimizes the use of available resources. For example, active learning can leverage the LLMs' retrieval quality and replace human feedback with LLM feedback in pseudo-relevance feedback scenarios.
LLMs have been used as judges, with active learning utilized to reduce the cost of annotation. For example, Zhang et al. \cite{zhang2023llmaaa} demonstrated that LLMs can act as annotators within an active learning framework, where the model selectively identifies informative samples for annotation. Their approach showed that task-specific models trained on only a few hundred LLM-generated labels can outperform the teacher LLM, significantly reducing annotation costs.
Xiao et al. \cite{xiao2023freeal} introduced a collaborative active learning framework that leverages LLMs as weak annotators to reduce human labeling requirements. Their approach combines model predictions with LLM-generated labels, enabling efficient training with fewer manually annotated examples.
Similarly, we leverage selective LLM feedback through active learning to reduce the cost of LLM-based relevance estimation, but we use it for reranking, rather than annotation.

\subsection{Long vs. Short Queries in Retrieval Models}
Search engines face challenges in interpreting long queries because the terms within a long query do not contribute equally to the user's information need \cite{gupta2015information}. As a result, users often submit short queries \cite{jansen2000real}, which provide limited contextual information and make it more difficult for retrieval systems to infer the underlying search intent.
For instance, an analysis of the MSN search log, which sampled roughly 15 million queries over a month, revealed an average query length of 2.4 words, with 90.3\% being less than five words long and 99.9\% being less than 12 words \cite{bendersky2009analysis}. While most queries are typically very short, Bendersky and Croft also observed an increase in the average query length over the past few years.

Throughout the years, short queries retrieval typically outperformed their longer counterparts when processed by the same information retrieval (IR) model.
Recognizing their potential, recent research has focused on developing techniques to enhance search effectiveness for long queries. As an example, Yan et al. \cite{yan2012hidden} investigated long queries that contain sentence-level structure rather than isolated keywords and proposed an HMM-based approach that exploits this structure to estimate the importance of different query terms and improve query representation.
Kumaran and Carvalho \cite{kumaran2009reducing} employ learning techniques to rank subsets of the original long query using quality prediction metrics.
In addition, Bendersky and Croft \cite{bendersky2008discovering} show that extracting key concepts from long queries can improve ranking performance.

\subsection{Budget-constrained LLM retrieval}

A growing line of work treats the number of LLM calls as an explicit, controllable resource rather than an afterthought, echoing the framing we adopt for LAAM. Kim et al. \cite{kim2026} propose BAGEL, which is applied to passages and models the relevance distribution over an entire embedding space with a query-specific Gaussian Process fit to sparse LLM relevance scores, then iteratively chooses which passages to submit to the LLM by balancing exploitation of high-confidence regions against exploration of uncertain ones. Huang et al. \cite{huang2026} take a related but distinct angle with TS-SetRank, arguing that a document's relevance is not fixed but context-dependent, varying according to which other candidates it is batched with and how they are ordered, and use uncertainty-aware sampling to estimate this ``contextual relevance'' efficiently rather than scoring every batch composition exhaustively. Both frameworks, like LAAM, replace exhaustive LLM evaluation with a model-guided selection policy, but they target the sampling/ordering of what is shown to the LLM rather than learning a surrogate that predicts the LLM's judgment for unlabeled documents, as LAAM does.

Other work reduces LLM cost by shrinking what is sent to the LLM rather than how many documents are sent. Li et al. \cite{li2024} propose KeyB2, which selects only the most informative blocks of a long document for LLM scoring instead of the whole text. Otero and Parapar \cite{otero2026} instead constrain the LLM's role within a classical pseudo-relevance feedback pipeline and use the LLM as a lightweight filter over the initial top-$k$ documents, retaining documents it judges relevant before estimating an RM3 expansion model from the filtered set. These approaches illustrate several complementary strategies for controlling LLM cost in retrieval. Specifically, they select which documents a priori to evaluate, how much of each document to evaluate, and what role the LLM plays within a larger pipeline. None learn to approximate the LLM using a lightweight model, as LAAM does.

\section{Model Architecture}
\begin{figure}[htbp]
    \centering
    \includegraphics[width=0.75\linewidth]{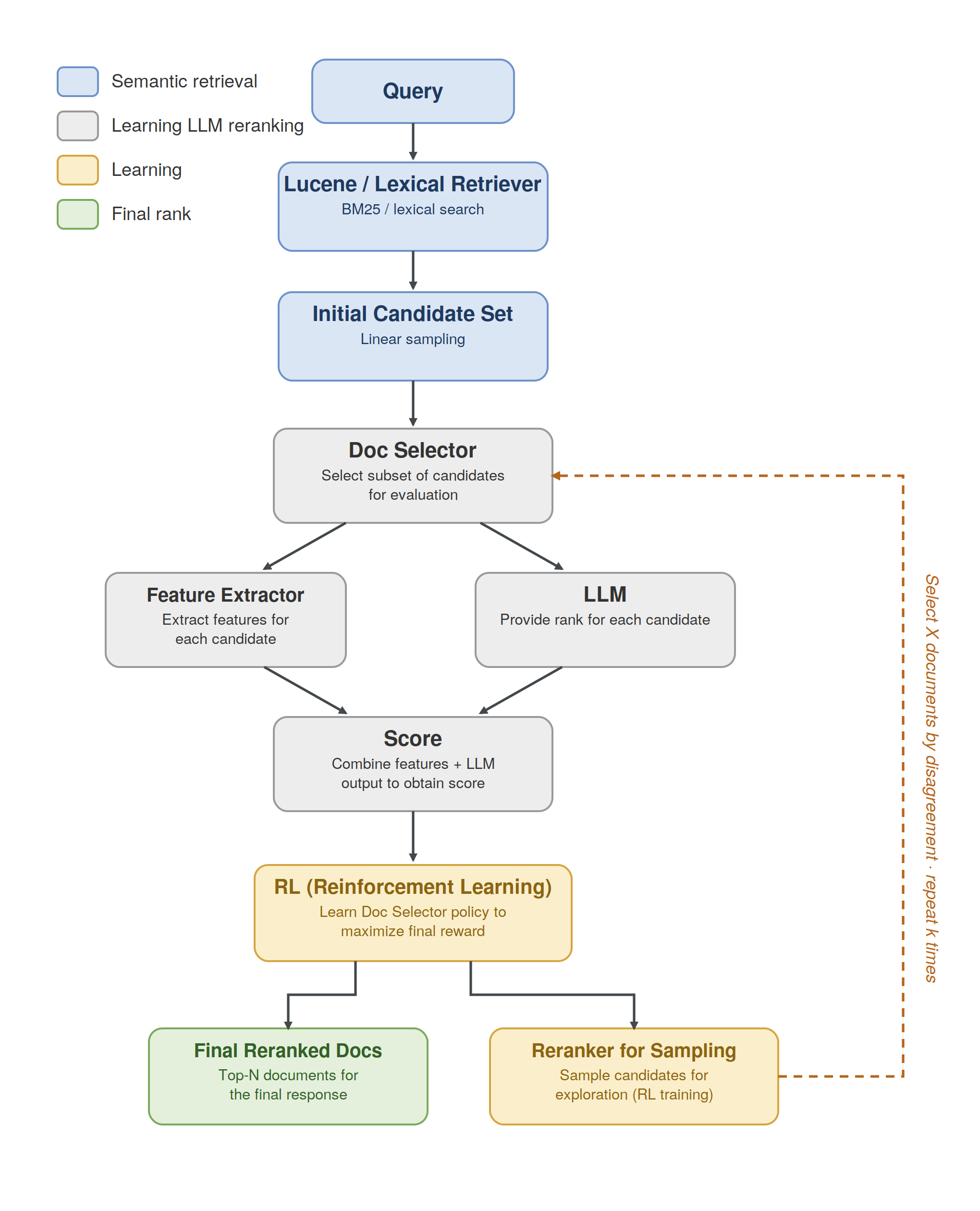}
    \caption{Architecture of the LAAM-based reranking retrieval pipeline.}
    \label{fig:pipeline}
\end{figure}

This research proposes a document retrieval framework that combines traditional search engines with Large Language Models (LLMs) to improve long-query retrieval while controlling computational cost. The framework learns to rerank documents that were ranked using BM25 using a small number of LLM-ranked documents. The system reduces the reliance on the LLM while preserving high retrieval quality by iteratively refining the LAAM. The hypothesis underlying the proposed model is that it is possible to take advantage of the strengths of the individual systems without succumbing to their drawbacks. Specifically, BM25 requires low computational cost but has lower accuracy, while the LLMs are a superior ranking function, provided at significant computational (and financial) cost. In the following we explain the proposed model, with reference to the flowchart in Figure~\ref{fig:pipeline}.

\paragraph{Lexical Search.}
The framework begins by processing a (long) query through a traditional search engine (Lexical retriever block in Figure~\ref{fig:pipeline}). Here we use Lucene with BM25 scoring, though other engines and scoring functions are possible. The search engine ranks the document corpus based on keyword relevance, producing an initial ranking with associated relevance scores.

\paragraph{LLM Scoring.}
The framework leverages an LLM to generate a high-quality relevance score for a subset of the documents returned by lexical retriever. In the first iteration we select an initial training set according to the methods described below, so as to ensure coverage across the ranking range. These sampled documents are scored by the LLM and are then used to train the LAAM. These tasks are shown in the Doc Selector and LLM blocks of Figure~\ref{fig:pipeline}.

\paragraph{Feature Extraction and Representation.}
For each document--query pair, the system extracts a feature representation combining lexical and semantic signals. Lexical relevance is captured by (i) the cosine similarity between the TF--IDF vectors of the query and the document, computed over a shared vocabulary of unigrams and bigrams with English stop-words removed, and (ii) the query-term overlap ratio (the fraction of query terms occurring in the document). Semantic relevance is captured by dense sentence-transformer embeddings of the query and document. This block is denoted as Feature Extractor in Figure~\ref{fig:pipeline}.

\paragraph{LLM Active Approximation Model (LAAM).}
The core of the proposed model is LAAM, an active-learning-based approximation framework that reduces the cost of LLM-based relevance assessment. LAAM learns to approximate LLM relevance judgments via a model applied to the feature representation described above, thereby reducing the need for costly LLM evaluations of all documents. We selected a Gradient Boosting regression model \cite{friedman2001greedy} for LAAM. The choice of Gradient Boosting was motivated by empirical evaluation against a Random Forest (RF) regressor \cite{breiman2001random}, where Gradient Boosting demonstrated superior predictive accuracy.

\paragraph{Greedy Active Learning Selection and Score Correction.}
Following scoring of the initial document set, the system enters an iterative active learning refinement phase. For each unevaluated (by the LLM) document, the LAAM estimates its LLM score (on a range of 1--1000), which is then linearly scaled into the scoring range of the lexical retrieval. The system calculates the disagreement (absolute difference) between the rescaled predicted score and the original search engine score for all documents and selects the single document with the highest disagreement between the predicted and actual lexical retrieval score in each iteration (the Reranker for Sampling block in Figure~\ref{fig:pipeline}). This pure exploitation strategy prioritizes documents where LAAM most strongly disagrees with the search engine ranking. The selected document is then evaluated by the LLM, and is added to the LAAM training set. Note that we also test an approach balancing exploration and exploitation.

We prioritize documents exhibiting the highest predicted disagreement between the LLM and lexical retrieval scores to ensure LAAM receives training labels that target areas of maximum discrepancy between the two models. The rationale for this is to prioritize documents which are most likely to benefit from LLM-based ranking. Other active learning acquisition strategies, such as uncertainty sampling based on score variance, remain promising avenues for future work.

\paragraph{Incremental Model Refinement.}
After each LLM evaluation, the LAAM is fine-tuned on the newly labeled example using incremental learning. This continuous refinement improves prediction accuracy over time and reduces future disagreements.

\paragraph{Convergence and Final Ranking.}
The iterative active learning cycle, consisting of greedy selection, LLM evaluation, score adjustment, and model refinement, continues until a convergence criterion is met. This convergence can be reached when discrepancies between predicted and actual rankings fall below a predefined threshold or the maximum number of allowed scored LLM documents is reached.

The final ranking is produced by sorting the documents by their LAAM-predicted scores. 

\section{Experimental Setup}
\subsection{Data}
We utilize the dataset from the TREC conference \cite{voorhees2004overview} Robust 2004 track. We focused on queries (a.k.a.\ topics) number 301--350. Each topic consists of a Title (short keyword-based query), Description (medium-length queries, clarifying user intent), and Narrative (long queries, defining relevance criteria). 

\subsection{Methodology}
We implemented LAAM with Apache Lucene~\cite{lucene} as the search engine and DeepSeek V3~\cite{liu2024deepseek} as the LLM. All experiments used topics 301--350 of the TREC ad-hoc track. We report Mean Average Precision (MAP) and precision at 1 and 10 (P@1 and P@10).

To compare systems, we scored the TREC-labeled (QRELs) documents among the first 3,000 documents retrieved by Lucene. The prompt to DeepSeek which was used to compute relevance is provided in the Appendix.

\subsubsection{Initial sampling strategy}
\label{sec:initial-sampling}
The composition of the initial labeled set determines the starting point of the active learning process before any prediction model has been trained. We evaluate six positional sampling strategies, each selecting the same budget of $15$ documents, from predetermined rank positions in the initial retrieval list. The initial budget size was arbitrarily chosen and kept fixed in all strategies to ensure a controlled comparison. The strategies differ only in how the sampling budget is distributed across the ranking:

\paragraph{Uniform Spread.} Samples at equal rank intervals. In our implementation the $p_i$ sampled document was chosen as $p_i = 1 + s i$ with step $s \approx 33$ for a sampling of 15 documents among the 500 top ranked documents.

\paragraph{Hybrid Sparse.} Balances sampling between highly ranked and lower-ranked documents through a quadratic rule $p_i = a\,i^{2} + b\,i + c$ with $(a,b,c) \approx (2.30,\,1.90,\,2.63)$.

\paragraph{Power-Law Family (Logarithmic Growth, Dense Top-Heavy, and Aggressive Top-50).} A family of top-emphasizing strategies sharing the power-law form $p_k = a\,k^{\beta}$, where the exponent $\beta$ acts as a single knob controlling how strongly sampling concentrates toward the top of the ranking: smaller $\beta$ yields a more top-heavy distribution. \emph{Logarithmic Growth} ($\beta \approx 2.78$) spreads positions across the full range, with $a = R / N^{\beta}$ so that the final position $p_N$ reaches the maximal rank $R$; \emph{Dense Top-Heavy} ($\beta = 2.25$, $a = 0.30$) focuses on the top $100$ ranks; \emph{Aggressive Top-50} ($\beta = 1.98$, $a = 0.30$) pushes this further, confining a near-quadratic core to the top $50$ ranks.

\paragraph{Fibonacci-Like.} Uses a golden-ratio progression as an alternative non-uniform spacing strategy, following the Fibonacci sequence $p_k = p_{k-1} + p_{k-2}$ ($1, 2, 3, 5, 8, 13, \dots$), whose consecutive ratios converge to the golden ratio $\varphi \approx 1.618$.

Parameters of each sampling strategy were chosen arbitrarily. Each strategy places its sampled positions according to a distinct spacing rule over the rank range $[1, 500]$. In all cases the computed values are rounded to the nearest integer rank. Furthermore, each strategy is evaluated using the proposed framework by measuring the convergence of P@10 as a function of the cumulative number of LLM evaluations, averaged over $50$ TREC queries. This experiment isolates the effect of the initial sampling distribution while keeping the retrieval pipeline, acquisition strategy, and labeling budget fixed.

\subsubsection{Document selection policy}
Three policies were tested for selecting which unevaluated documents to present to the LLM at each iteration: (1) a greedy policy that always selects the document with maximal disagreement between the predicted and actual lexical score, (2) a random uniform policy serving as an uninformed baseline, and (3) an inverse greedy policy that selects the document with minimal disagreement (i.e., those that are least uncertain under the current predictor). (4) exploration-exploitation balance, using either $\varepsilon$-greedy (with $\varepsilon=0.3$ providing the best results) or Boltzmann sampling (with $T=0.5$ providing the best results).

\subsubsection{Embedding model selection}
We compared three pre-trained sentence encoder models differing in dimensionality and training objective: all-MiniLM-L6-v2 (384 dimensions, optimized for speed), BGE-base-en-v1.5 (768 dimensions, trained explicitly for retrieval tasks on BEIR/TREC benchmarks), and GTE-base (768 dimensions, general-purpose).

\subsubsection{Feature representation}
To quantify the contribution of each feature group to LAAM prediction performance, we performed a leave-one-group-out ablation study of four feature configurations: the full 770-dimensional representation comprising semantic query--document embeddings, TF-IDF cosine similarity, and term-overlap features, and three reduced variants obtained by independently removing TF-IDF similarity, term overlap, or semantic embeddings. The relative contribution of the retained feature groups was further analyzed using linear regression over per-query P@10 scores.

\subsubsection{LAAM and its hyperparameters}
The LAAM is used to predict LLM-derived relevance scores for documents from their extracted features. In preliminary experiments, a Random Forest Regression predictor showed worse results than Gradient Boosting Regression, and we therefore focused on Gradient Boosting. The Gradient Boosting model is retrained whenever new labels from the LLM are added. 

We provide results from the default settings of the Gradient Boosting implementation, but show sensitivity to its parameters by performing a systematic grid search over Gradient Boosting hyperparameters across 64 configurations defined by four values for the number of estimators (\{10, 50, 300, 1500\}), maximum tree depths (\{1, 4, 15, 60\}), and learning rates ($10^{-5}, 5 \times 10^{-4}, 5 \times 10^{-3}, 5 \times 10^{-2}$).

\subsubsection{Energy Efficiency}
We measured the energy consumption of the lexical retriever and our hybrid framework directly at the hardware level using the Linux \texttt{perf} subsystem. The measurements of our system included the full execution pipeline, including BM25 retrieval, feature extraction, embedding generation, LAAM prediction, active learning selection, and all associated preprocessing overhead.
Direct hardware level measurement of DeepSeek was not possible because inference was performed through a remote API service. Consequently, its energy consumption was estimated using the infrastructure-aware benchmarking methodology proposed by Jegham et al.~\cite{jegham2025hungry}, which models LLM inference energy based on public API performance statistics, hardware inference estimates, and data-center-level efficiency factors.

\section{Results}

\subsection{End-to-End Retrieval Performance}
Table~\ref{tab:main_results} shows retrieval accuracy for the four evaluated systems: the BM25-Lucene retrieval, full DeepSeek ranking of the complete candidate set, and LAAM after $K=50$ and $K=100$ iterations.

\begin{table}[htbp]
\centering
\caption{Retrieval accuracy by system and query type. All nine settings (3 query verbosity levels of 3 performance measures) showed significant differences (Friedman test, $p<0.05$, with Bonferroni correction).}
\label{tab:main_results}
\renewcommand{\arraystretch}{1.15}
\small
\begin{tabular}{l|ccc|ccc|ccc}
\toprule
\textbf{Method} & \multicolumn{3}{c|}{\textbf{Narrative}} & \multicolumn{3}{c|}{\textbf{Description}} & \multicolumn{3}{c}{\textbf{Title}} \\
\cmidrule{2-10} 
 & P@1 & P@10 & MAP & P@1 & P@10 & MAP & P@1 & P@10 & MAP \\
\midrule
Lucene (BM25)
& 0.490 & 0.243 & 0.187
& 0.360 & 0.205 & 0.185
& 0.520 & 0.353 & 0.287 \\
DeepSeek (full)
& 0.674 & 0.526 & 0.449
& 0.180 & 0.128 & 0.157
& 0.200 & 0.108 & 0.136 \\
\textbf{LAAM ($K=50$)}
& 0.771 & 0.412 & 0.230
& 0.360 & 0.141 & 0.151
& 0.520 & 0.284 & 0.246 \\
\textbf{LAAM ($K=100$)}
& 0.735 & 0.459 & 0.257
& 0.380 & 0.169 & 0.151
& 0.520 & 0.305 & 0.237 \\
\midrule
p-value & 0.001 & $<0.001$ & $<0.001$ & $<0.001$ & $<0.001$ & $<0.001$ & 0.020 & 0.001  & $<0.001$ \\
\bottomrule
\end{tabular}
\end{table}

As shown in Table~\ref{tab:main_results}, retrieval effectiveness varies substantially across query types. Narrative queries benefit most from the proposed method. On this query type LAAM outperforms BM25 across all three measures, and the gap with exhaustive DeepSeek reranking narrows as the labeling budget increases. These gains are particularly notable given that LAAM requires LLM evaluations for only a small subset of candidate documents. To assess statistical significance, we applied the Friedman test to the per-query scores of the four systems for each measure and query type. All nine settings showed significant differences ($p<0.05$, with Bonferroni correction). Following this analysis, we performed post-hoc pairwise Wilcoxon signed-rank tests with Holm correction. These tests confirmed that LAAM significantly outperforms the Lucene baseline on P@10 and MAP across all query types, as well as on P@1 for narrative and description queries. LAAM is statistically indistinguishable from exhaustive DeepSeek reranking in P@1 and P@10 for narrative and description queries, though DeepSeek is statistically significantly better than LAAM in MAP.

\subsection{Energy Efficiency}
\begin{figure}[htbp]
    \centering
    \includegraphics[width=0.75\linewidth]{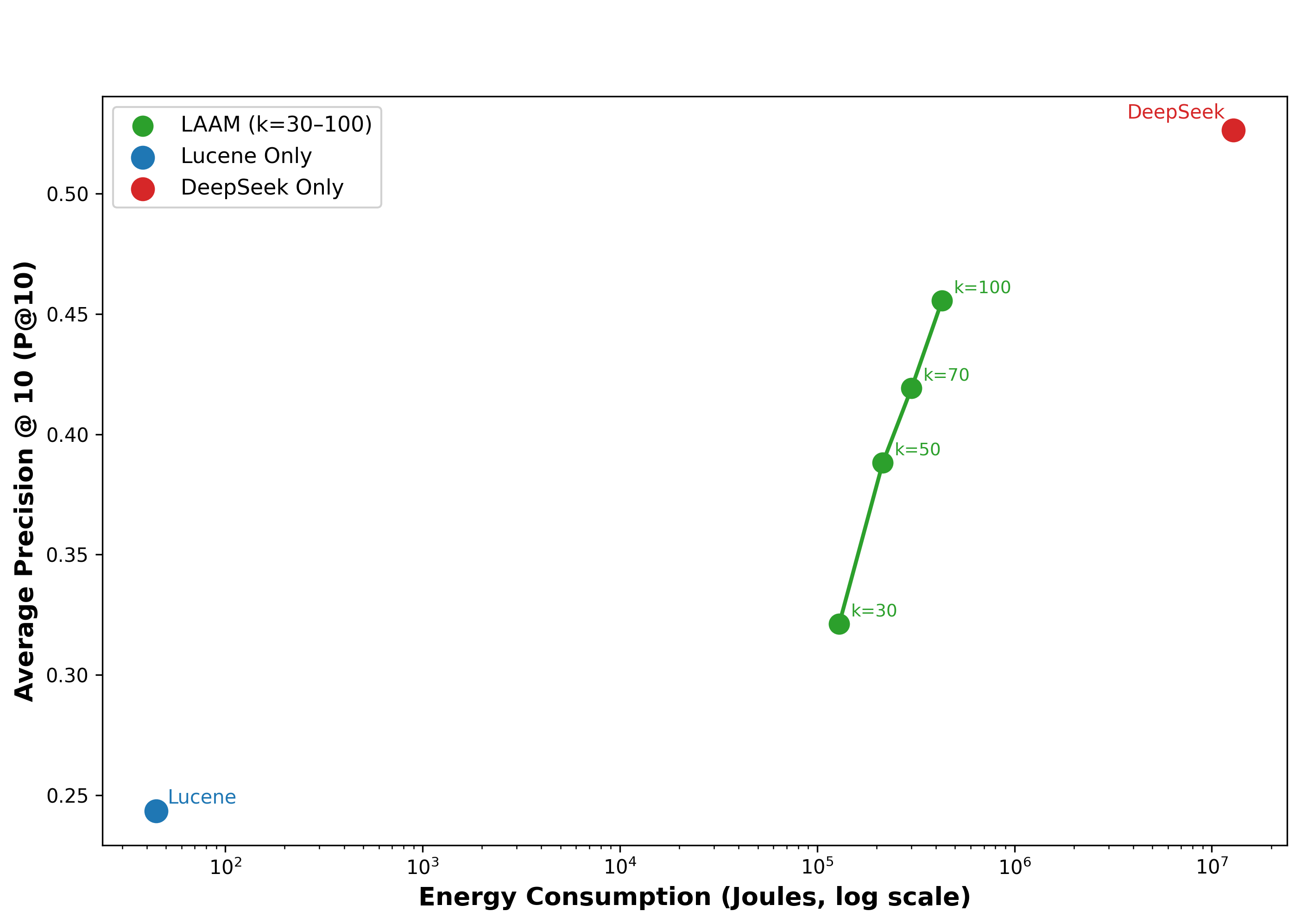}
    \caption{Energy--accuracy trade-off. Average P@10 on narrative queries plotted against per-query energy consumption on a logarithmic scale. LAAM is plotted with different document labeling budgets.}
    \label{fig:energy_tradeoff}
\end{figure}

Figure~\ref{fig:energy_tradeoff} shows the trade-off between retrieval quality and energy consumption. As Figure~\ref{fig:energy_tradeoff} demonstrates, there is an exponential relationship between retrieval accuracy and computational cost. Under the $K=50$ evaluation budget, the proposed framework consumes only $1.7\%$ of the energy required for full DeepSeek reranking ($215~\mathrm{kJ}$ versus $12.9~\mathrm{MJ}$), while reaching approximately $74\%$ of its average P@10 accuracy ($0.388$ versus $0.525$). Increasing the evaluation budget to $K=100$ raises energy consumption to $3.3\%$ of DeepSeek's cost ($430~\mathrm{kJ}$), while improving average P@10 to $0.456$, corresponding to roughly $87\%$ of DeepSeek's ranking accuracy.

Two observations are particularly important. First, as Figure~\ref{fig:energy_tradeoff} shows, the proposed hybrid retrieval framework achieves a highly favorable energy to accuracy ratio in the low-budget regime, indicating that a relatively small number of selectively chosen LLM evaluations captures most of the achievable retrieval gains. Second, as can be seen in Figure~\ref{fig:energy_tradeoff}, the improvement between $K=50$ and $K=100$ exhibits diminishing returns, as doubling the number of LLM evaluations doubles the energy consumption, but yields only a modest increase in retrieval quality. This saturation behavior suggests that the majority of high value ranking corrections are concentrated within the first LLM evaluations selected by the initial active learning sampling policy.

\subsection{Initial Sampling Position Strategy}
\begin{figure}[htbp]
    \centering
    \includegraphics[width=0.75\linewidth]{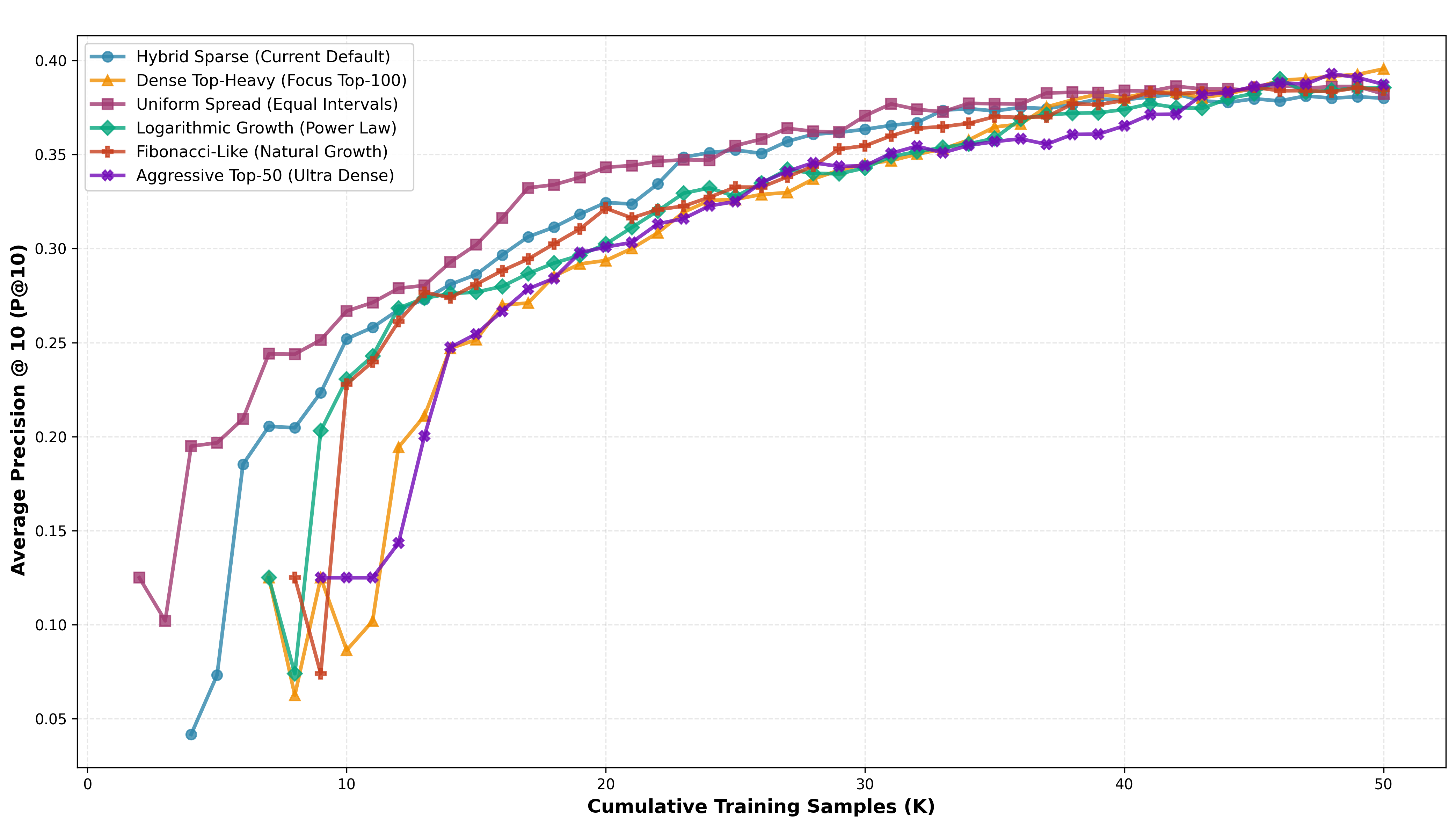}
    \caption{Average P@10 versus the number of LLM evaluations $K$ for different initial sampling strategies. While starting points differ, all converge after approximately 50 documents are labeled.}
    \label{fig:position_comparison}
\end{figure}

Figure~\ref{fig:position_comparison} presents P@10 as a function of the number of LLM evaluations $K$ for the six positional sampling strategies. All six sampling strategies converge to similar P@10 values by approximately $K=50$. In the cold-start phase ($K<15$), the top-concentrated Aggressive Top-50 and Dense Top-Heavy strategies exhibit greater instability, as sampling predominantly from the highest ranks provides limited exposure to non-relevant documents. In contrast, broader sampling strategies, such as Uniform Spread and Hybrid Sparse, produce more stable early convergence by exposing the model to a more diverse set of relevance signals. Despite this initial disadvantage, Dense Top-Heavy converges more rapidly and achieves the highest final performance at $K=50$ (average $\text{P@10}=0.40$). By focusing its initial samples on the highest-ranked documents, it allocates supervision to the region most directly affecting P@10, allowing subsequent LLM evaluations to improve the head of the ranking more efficiently. Overall, these results highlight a trade-off between early training stability and final retrieval effectiveness: broader sampling improves initialization, whereas top-concentrated sampling ultimately yields superior retrieval performance.

\subsection{Document Features}
\begin{figure}[htbp]
    \centering
    \includegraphics[width=0.75\linewidth]{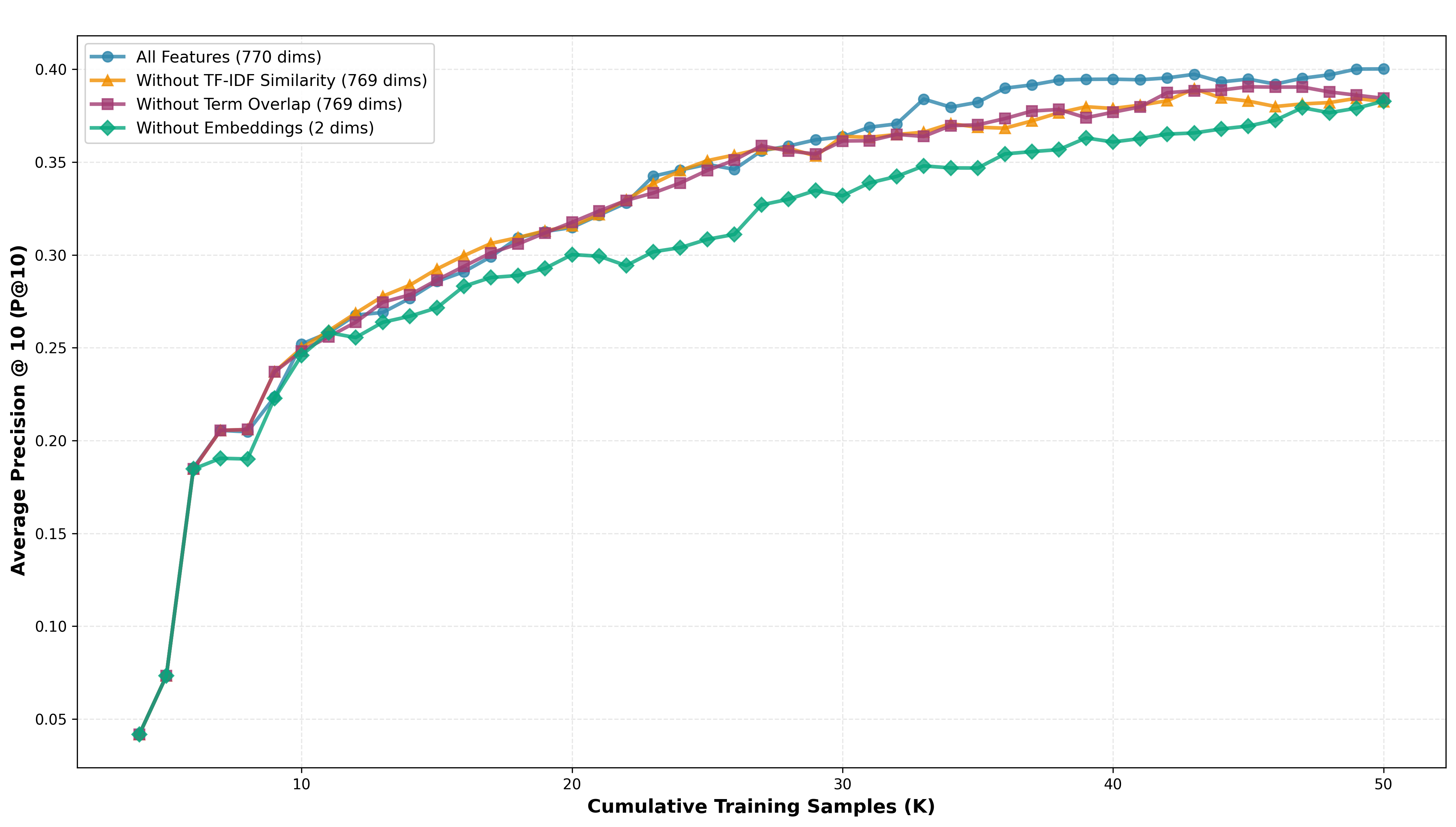}
    \caption{Ablation results showing average P@10 versus the cumulative number of LLM evaluations $K$ for different feature families.}
    \label{fig:feature_comparison}
\end{figure}

Figure~\ref{fig:feature_comparison} presents the average P@10 as a function of the cumulative number of LLM evaluations $K$ for four feature configurations: the complete 770 dimensional representation and three leave-one-group-out ablations.

The complete feature representation reaches a slightly higher average P@10. The largest performance degradation is observed when embedding features are removed. By $K=50$, the gap between the full feature configuration and the variant \textit{Without Embeddings} reaches near $0.04$ average P@10 points. Removing either the similarity features or lexical term-overlap features results in smaller but still measurable declines in retrieval quality. These results indicate that dense embedding representations provide the dominant contribution to ranking quality, while sparse lexical signals supply complementary improvements that further stabilize retrieval performance in both low and moderate sample scenarios.

A linear regression model of the per-query P@10 as a function of the inclusion or exclusion of each feature family reaches $R^2 = 0.414$. The slope of document embeddings is largest ($\beta=0.0168$), followed by term overlap ($\beta=0.0068$) and query embeddings ($\beta=0.0061$). TF-IDF similarity has the smallest slope ($\beta=0.0004$). This indicates that dense semantic representations capture most of the signal required to approximate LLM relevance judgments, while lexical and query-level features contribute more modestly but remain complementary. These results help explain LAAM's strong ranking performance despite substantially reducing direct LLM evaluations, as further discussed in Section~\ref{sec:discussion}.

\subsection{Embedding Models}
\begin{figure}[htbp]
    \centering
    \includegraphics[width=0.75\linewidth]{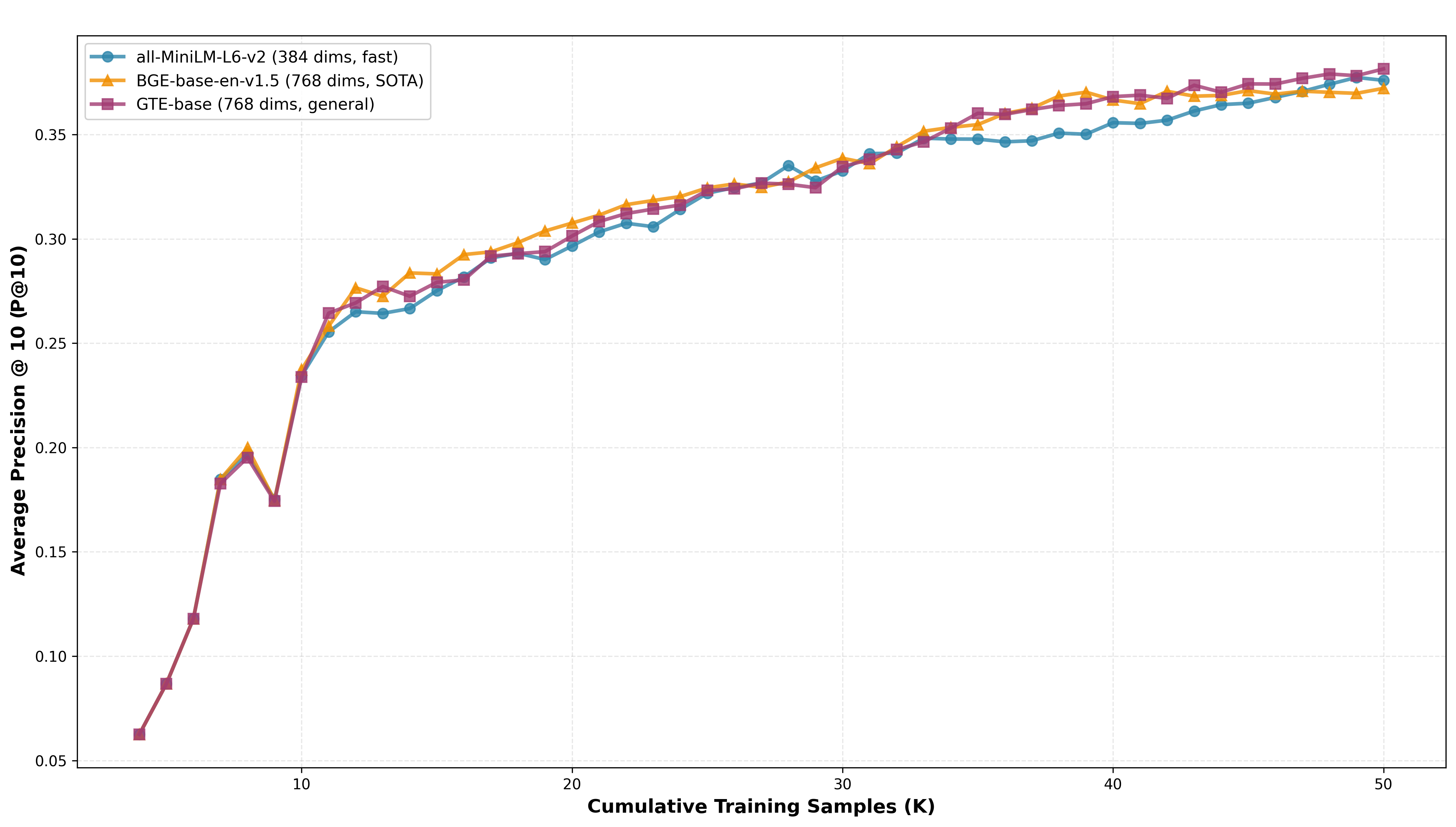}
    \caption{P@10 versus the number of LLM evaluations $K$ for different embedding models.}
    \label{fig:embedding_comparison}
\end{figure}

Figure~\ref{fig:embedding_comparison} compares P@10 with cumulative training size $K$ for three sentence encoder models with differing embedding dimensionalities. The two 768 dimensional encoder models, \texttt{BGE-base-en-v1.5} and \texttt{GTE-base}, achieve nearly identical performance across the full evaluation range, converging to P@10 values of near $0.40$ at $K=50$. Although the 384 dimensional \texttt{all-MiniLM-L6-v2} encoder consistently trails the larger models by roughly $0.02$ P@10, its markedly lower dimensionality offers important practical advantages, including reduced memory consumption, faster embedding computation, and lower retrieval latency. Given the relatively modest performance gap, the MiniLM model therefore provides a favorable efficiency-effectiveness trade-off for deployment scenarios where computational cost and scalability are primary constraints.

\subsection{LAAM Selection Strategy}
\begin{figure}[htbp]
    \centering
    \includegraphics[width=0.75\linewidth]{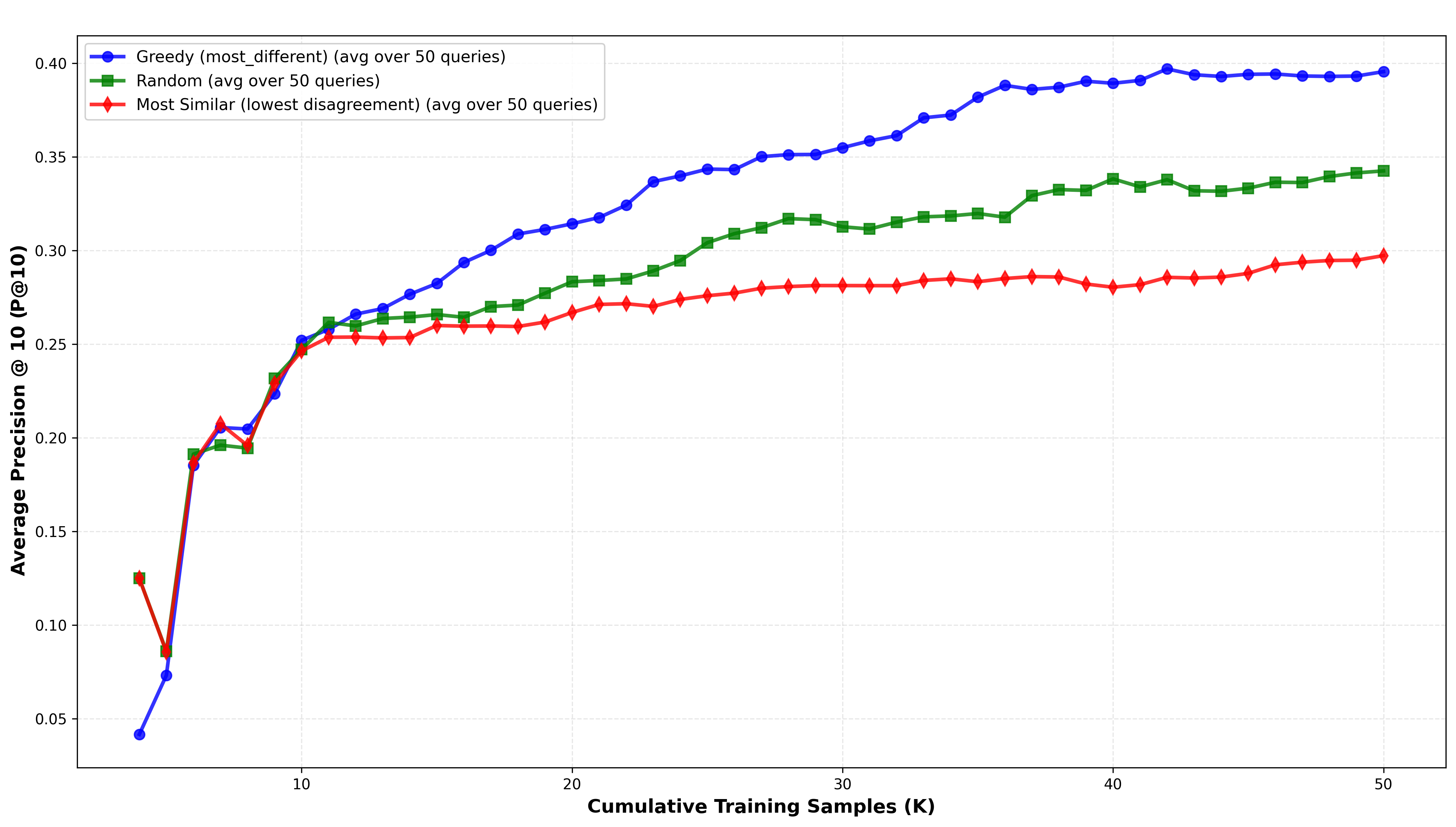}
    \caption{Average P@10 as a function of $K$ for three LAAM selection strategies. The greedy strategy consistently outperforms both the random and the most similar selection throughout the LAAM process.}
    \label{fig:selection_strategy}
\end{figure}

Figure~\ref{fig:selection_strategy} compares average P@10 for three document selection policies applied at each LAAM iteration: \textit{Greedy} (maximum disagreement), \textit{Random} (uniform sampling), and \textit{Most Similar} (minimum disagreement).
As shown in Figure~\ref{fig:selection_strategy}, the \textit{Greedy} strategy consistently outperforms both \textit{Random} and \textit{Most Similar} selection throughout the LAAM process. Thus, disagreement-based selection provides substantially higher information gain per annotation compared to similarity-based or random sampling strategies.

\subsection{LAAM Convergence}
\begin{figure}[htbp]
    \centering
    \includegraphics[width=0.75\linewidth]{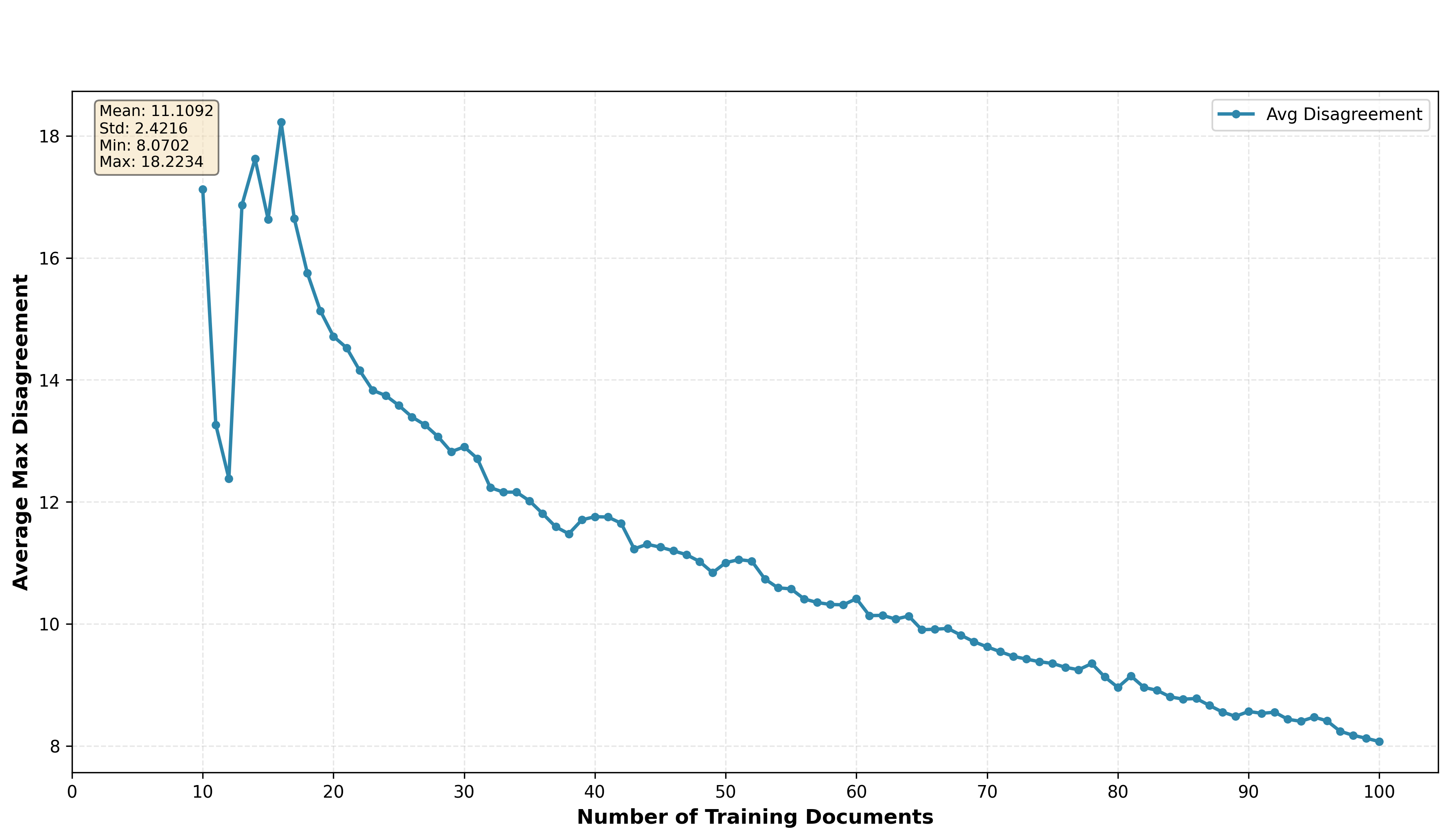}
    \caption{Average maximum disagreement as a function of LLM evaluations size $K$, averaged over all queries.}
    \label{fig:active_learning_convergence}
\end{figure}

Figure~\ref{fig:active_learning_convergence} shows the evolution of the average maximum disagreement across all the 50 queries as a function of the cumulative number of LLM evaluations. For this analysis, we report results using the best-performing configuration of our method. Disagreement is evaluated as:
\[
\Delta_{\max} = \max_{d \notin L} \left| \hat{s}_{\theta}(d) - s_{L}(d) \right|,
\]
where $\hat{s}_{\theta}(d)$ denotes the LAAM score and $s_{L}(d)$ represents the lexical retrieval score for document $d$.

As illustrated in Figure~\ref{fig:active_learning_convergence}, the disagreement curve exhibits a brief initial increase, followed by an almost monotone decrease as additional LLM evaluations are incorporated.
The early peak confirms that the predictor rapidly acquires a sufficient signal to identify highly misranked or uncertain documents during the initial stages of LAAM. Subsequently, the monotonic decline indicates progressive alignment between the learned predictor and the lexical retrieval as the high uncertainty samples are resolved. The asymptotic floor likely reflects residual disagreement that cannot be eliminated through additional sampling, corresponding to intrinsic differences between the learned representation and the lexical retrieval model.

\subsection{Balancing exploration with exploitation}

Table~\ref{tab:exploration_exploitation} shows a comparison of the narrative retrieval results with LAAM approaches that balance exploration with exploitation. The results are for the best performing values of $\epsilon$ and $T$, compared to the fully exploitive approach. As the table demonstrates, the exploitation approach (selecting the document with the maximal difference between predicted and actual lexical retrieval scores) provides the best results.

\begin{table}[htbp]
\centering
\caption{Results when balancing exploration with exploitation. The table shows the results of the narrative queries after 50 iterations.}
\label{tab:exploration_exploitation}
\begin{tabular}{lccc}
\toprule
\textbf{Method} & \textbf{P@1} & \textbf{P@10} & \textbf{MAP} \\
\midrule
Exploitation-only & 0.771 & 0.412 & 0.230 \\
$\epsilon$-Greedy ($\epsilon=0.3$) & 0.700 & 0.382 & 0.214 \\
Boltzmann sampling ($T=0.5$) & 0.660 & 0.376 & 0.216 \\
\bottomrule
\end{tabular}
\end{table}

\section{Discussion}
\label{sec:discussion}

The rapid adoption of LLMs has changed information retrieval by enabling systems to interpret complex, natural-language information needs and go beyond simple keyword matching. However, these gains come at a substantial computational and energy cost, since LLM-based relevance estimation can be several orders of magnitude more expensive than traditional lexical retrieval methods \cite{scells2022reduce}. Existing retrieval pipelines therefore face a difficult trade-off: relying solely on lexical methods sacrifices semantic understanding, whereas reranking retrieved documents with an LLM is expensive. From this perspective, retrieval becomes a resource-allocation problem whereby, given a limited computational budget, which documents are worth spending an expensive LLM evaluation on, and which can be ranked adequately using inexpensive lexical methods? The proposed framework addresses this problem by learning to approximate LLM relevance judgments and allocating LLM evaluations only to documents where they are expected to provide the greatest benefit. The resulting system approaches the accuracy of exhaustive LLM reranking while consuming only a small fraction of the computational and energy cost.

Our central finding is that selective LLM evaluation, guided by a disagreement-based LAAM policy, recovers most of the retrieval quality of exhaustive LLM reranking at a fraction of its computational cost and in some cases surpasses it. For example, on narrative queries the framework reaches $\mathrm{P}@10 = 0.412$ at $K = 50$ (78\% of the exhaustive $\mathrm{P}@10 = 0.526$) and $\mathrm{P}@10 = 0.459$ at $K = 100$ (approximately 87\%), while consuming only 1.7\% and 3.3\% of the per-query energy, respectively.
We attribute this to the hybrid system's ability to combine the lexical prior provided by BM25 with targeted semantic corrections from the LLM. Selective evaluation concentrates the computational effort on the documents most likely to be ranked dissimilarly to LLM rankings, capturing the majority of the gains available from exhaustive reranking while avoiding the cost of evaluating every candidate. The resulting performance demonstrates that a relatively small number of strategically chosen LLM evaluations is sufficient to recover most of the improvement obtainable from full semantic reranking.

Performance differences across query types highlight an important limitation of purely lexical retrieval methods. For short title and description queries, BM25 is often sufficient because the information need is expressed through a small set of keywords. In these cases, even advanced semantic models have limited additional context from which to infer relevance. In contrast, narrative queries provide detailed descriptions of the user's intent, including relevance criteria, exclusions, negations, and nuanced distinctions that cannot be captured through term matching alone. LLMs are particularly effective in this setting because they can interpret these semantic constraints and reason about document relevance \cite{weller2025followir}.
Our framework is designed to exploit the arbitrage between LLMs and semantic retrieval. When this difference is unavailable, the advantage of our proposed system diminishes.

We highlight the importance of evaluating retrieval effectiveness alongside energy consumption, as accuracy gains can come at substantial computational cost. The results reveal a clear energy--accuracy trade-off: increasing the LLM evaluation budget improves average P@10, but the improvements become progressively smaller relative to the additional energy required. This produces a non-linear curve characterized by diminishing returns rather than a proportional relationship between energy expenditure and retrieval quality. Our framework occupies an intermediate region between lexical retrieval and LLM reranking. While full reranking achieves the highest average P@10, it requires 1--2 orders of magnitude more energy than our approach.

These findings suggest that retrieval systems should adapt their evaluation budget to query complexity. Short keyword-based queries can be handled efficiently with lexical retrieval, whereas long queries justify a larger semantic reranking budget because they benefit substantially more from LLM-based relevance assessment. This flexibility is a key advantage of the proposed framework, as it enables the accuracy on long queries to be scaled according to a predefined computational budget. Rather than committing to either purely lexical retrieval or LLM reranking, the system provides a continuous spectrum of operating points, allowing practitioners to select the most appropriate balance between effectiveness and energy consumption for their deployment constraints.

Our work has several limitations. First, all experiments use DeepSeek \cite{liu2024deepseek}, chosen for its favorable accuracy--cost trade-off. Better reasoning models may achieve comparable accuracy with fewer LLM calls, potentially enabling earlier stopping within the active learning loop. Models that emphasize energy efficiency may reduce the gap in accuracy between the proposed system and full reranking.

TREC Robust 2004 is a public benchmark whose documents and relevance judgments are widely distributed online and are thus possibly included in the training corpora of contemporary LLMs. As a result, some of DeepSeek's relevance judgments may reflect memorized patterns rather than pure inference, potentially making absolute performance estimates optimistic compared to fully unseen corpora. This is a general limitation of LLM-based evaluation on widely available benchmarks. However, its impact is partially mitigated in our setting, as the framework does not consistently outperform BM25 on title and description query types, suggesting that any potential memorization does not lead to consistent improvements in every case.

While per-query energy consumption for BM25 and the hybrid framework is measured directly at the hardware level using the Linux \texttt{perf} tool, the same is not possible for the full DeepSeek reranking configuration. Due to the lack of local access to the underlying model infrastructure, DeepSeek inference was executed via a remote API, preventing direct measurement under identical hardware conditions. As a result, its energy consumption is estimated using the infrastructure-aware methodology of Jegham et al.~\cite{jegham2025hungry}. While this methodology provides a principled and reproducible estimate grounded in prior work, it cannot fully replicate direct hardware-level measurement under identical experimental conditions. Consequently, energy comparisons involving the exhaustive DeepSeek configuration should be interpreted as estimates rather than exact measurements, although the reported trends are expected to remain representative.

Several directions exist for future extensions of this work. First, in the current framework, the LAAM is trained independently for each query. The LAAM is intended to identify documents for which the BM25 ranking may not accurately reflect true relevance and where LLM evaluation is most likely to improve ranking quality. Transferring knowledge from previously processed queries could provide a stronger initialization, enabling the LAAM to recognize such cases more quickly and reducing the number of LLM evaluations required to reach a given level of retrieval quality.

Second, the dissimilar results across query types motivate an upstream classifier that routes short queries directly to BM25 and reserves the hybrid pipeline for long queries, translating the verbosity-sensitivity established above into a system efficiency gain.

Third, the current evaluation was limited to TREC Robust 2004. Testing the framework on newer or private collections that are less likely to overlap with LLM training data would provide stronger evidence that the observed gains generalize beyond this benchmark.

Finally, although this work focuses on document reranking, the proposed architecture could be applied within retrieval-augmented generation (RAG) systems. Using LAAM to reduce retrieval costs while maintaining retrieval quality may improve the overall efficiency of end-to-end RAG pipelines.


\clearpage
\bibliographystyle{plainnat}
\bibliography{references}

\clearpage
\appendix
\section{LLM Prompt}
\label{app:llm-prompt}

The following prompt was used to instruct the LLM to assign a relevance
score to each document with respect to a given query.

\begin{llmprompt}
\noindent \textbf{System Prompt}: You are a document's relevant ranking model.

\medskip
\noindent \textbf{User Prompt}: You are an expert ranking assistant. Your task: Given the following query and document, assign a relevance score to the document with respect to the query on a scale from 1 (completely irrelevant) to 1000 (perfectly relevant).

\medskip
\noindent \textbf{Instructions:}
\begin{itemize}[leftmargin=*, nosep]
    \item Output ONLY a single integer between 1 and 1000, followed by
    a brief explanation of your decision.
    
    \item Use the full range: choose precise, fine-grained numbers
    (e.g., 2, 16, 532, 876, 999) to reflect nuanced judgments.
    
    \item Avoid always using rounded or extreme values (like 1, 10,
    100, 500, 1000); instead, select diverse and realistic numbers
    that best represent the document's relevance.
    
    \item Do not repeat the same score for different examples unless
    their relevance is truly identical.
    
    \item Be as accurate and specific as possible in your scoring.
\end{itemize}

\medskip
\noindent \textbf{Query:} \{query\} \\
\noindent \textbf{Document:} \{doc\_content\} \\
\noindent \textbf{Relevance Score (1--1000):}
\end{llmprompt}


\end{document}